\documentclass[conference]{IEEEtran}
\IEEEoverridecommandlockouts
\usepackage[ruled,lined]{algorithm2e}
\usepackage{setspace}
\usepackage[ruled,lined]{algorithm2e}
\usepackage{balance}

\usepackage{amsthm}
\newtheorem{mydef}{Proposition}

\newtheorem{mydef5}{Result}
\newtheorem{mydef6}{Remark}
\usepackage{mathrsfs}  
\usepackage{graphicx}
\usepackage{xcolor}
\def\BibTeX{{\rm B\kern-.05em{\sc i\kern-.025em b}\kern-.08em
    T\kern-.1667em\lower.7ex\hbox{E}\kern-.125emX}}

\usepackage{graphicx,epsf,epic,eepic,psfig,latexcad}
\renewcommand{\baselinestretch}{1}

\usepackage{amssymb,amsmath,caption}
 \usepackage{cite}  

\usepackage{amsmath, amssymb, amsthm}
\usepackage{graphicx}
\renewcommand{\baselinestretch}{1}

\usepackage{epstopdf}

\DeclareMathOperator*{\argmin}{arg\,min} 
\newcommand\numeq[1]%
  {\stackrel{\scriptscriptstyle(\mkern-1.5mu#1\mkern-1.5mu)}{=}}
\makeatletter
\usepackage{textcomp}
\makeatother
\usepackage{xcolor}
\begin{document}
	\title{
Integrated Sensing and Communications for Unsourced Random Access: Fundamental Limits}

\author{
	\IEEEauthorblockN{Mohammad Javad Ahmadi$^{\star}$ and Rafael F. Schaefer$^{\star,\dagger}$\\[0.5ex]}
	\IEEEauthorblockA{\small $^{\star}$ Chair of Information Theory and Machine Learning,
		Technische Universit\"at Dresden\\
		$^{\star}$ Cluster of Excellence CeTI, 
            $^{\dagger}$ BMBF Research Hub 6G-life\\
            $^{\dagger}$ Barkhausen Institut, Dresden, Germany\\[0.5ex]
		email: \texttt{\{mohammad\_javad.ahmadi, rafael.schaefer\}@tu-dresden.de}
	}
	\and
	\IEEEauthorblockN{H. Vincent Poor\\[0.5ex]}
	\IEEEauthorblockA{\small Dept. of Electrical and Computer Engineering\\
		Princeton University\\[0.5ex]
		email: \texttt{poor@princeton.edu}
	}
	\thanks{This work of M. J. Ahmadi and R. F. Schaefer was supported by the German Research Foundation (DFG) as part of Germany’s Excellence Strategy - EXC 2050/1 - Project ID 390696704 - Cluster of Excellence \emph{``Centre for Tactile Internet with Human-in-the-Loop'' (CeTI)} of Technische Universität Dresden. This work of R. F. Schaefer was further supported in part by the German Federal Ministry of Education and Research (BMBF) within the national initiative on 6G Communication Systems through the research hub \emph{6G-life} under Grant 16KISK001K. This work of H. V. Poor was supported by the U.S. National Science Foundation under Grant ECCS-2335876.
}
}

\maketitle

\begin{abstract}
This work considers the problem of integrated sensing and communications (ISAC) with a massive number of unsourced and uncoordinated users. In the proposed model, known as the unsourced ISAC system (UNISAC), all active communication and sensing users simultaneously share a short frame to transmit their signals, without requiring scheduling with the base station (BS). Hence, the signal received from each user is affected by significant interference from numerous interfering users, making it challenging to extract the transmitted signals. UNISAC aims to decode the transmitted message sequences from communication users while simultaneously detecting active sensing users and estimating their angles of arrival, regardless of the identity of the senders. In this paper, we derive an approximate achievable result for UNISAC and demonstrate its superiority over conventional approaches such as ALOHA, time-division multiple access, treating interference as noise, and multiple signal classification. Through numerical simulations, we validate the effectiveness of UNISAC's sensing and communication capabilities for a large number of users.
\end{abstract}

\begin{figure}
	\centering
	\includegraphics[width=1\linewidth]{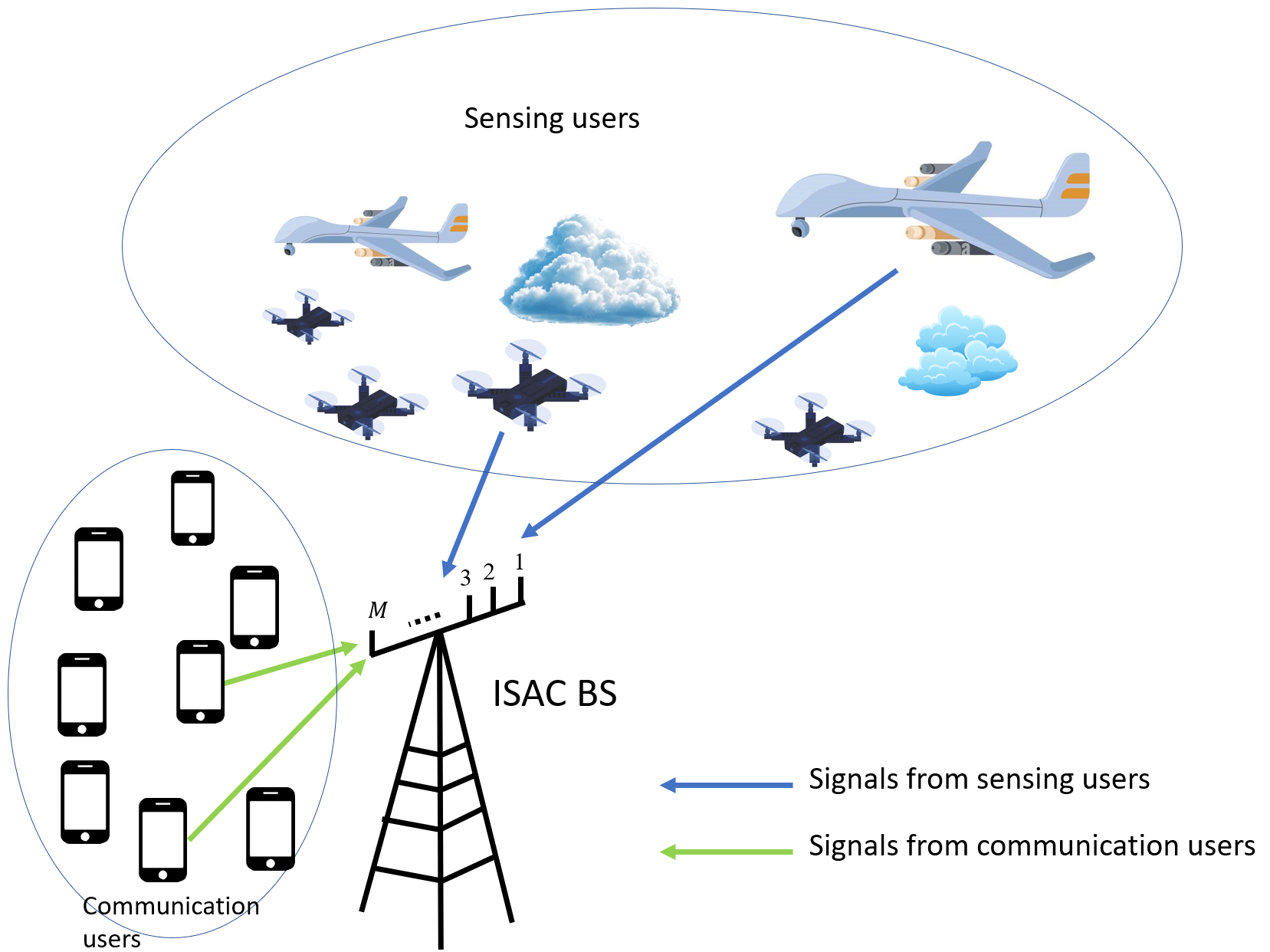}
	\caption{{\small Illustration of the UNISAC model. The system accommodates multiple users for sensing and communication purposes, with only a limited subset active at any given moment.}}
	\label{Fig_configuratin}
 \vspace{-4mm}
\end{figure}
\section{Introduction}
\label{introduction}
In unsourced random access (URA), introduced by Polyanskiy in \cite{polyanskiy2017perspective}, a massive number of unidentified users transmit sporadically over a short time period without any prior scheduling with the base station (BS). Moreover, the BS is only concerned with recovering the transmitted information, without identifying the sender (the unsourced feature). Eliminating the need for scheduling and ignoring user identities makes URA a promising solution for supporting an unbounded number of users. \textcolor{black}{This is important because user-specific scheduling and identification introduce significant signaling overhead, increase computational complexity and delay, and cause higher energy consumption, all of which are undesirable in large-scale networks.} This approach is particularly suitable for the next generation of communication networks, where millions of inexpensive devices are connected to the system, with only a fraction of them active during each transmission~\cite{andreev2020polar,Ozates2023Slotted,Gkagkos2023FASURA,Fengler2023Coded,glebov2019achievability,pradhan2020polar,polyanskiy2017perspective,ahmadi2021random,Zhang2025SparseCode,ahmadi2021Unsourced,Ahmadi2024RISUMA,Ahmadi2023RIS,Ahmadi2025Feedback,Ahmadi_PHD_thesis}. {\color{black}In such scenarios, traditional user-specific scheduling and identification become impractical and inefficient, reinforcing the relevance of URA for future massive access networks.}

Existing research on URA has predominantly concentrated on extracting information transmitted by unsourced users (communication purpose). However, with the advent of sixth-generation (6G) mobile networks, additional functionalities such as computing, sensing, and security have become imperative~\cite{Qi2022integrating,Xu2022Anti}. As hardware architectures, channel characteristics, and signal processing methods become increasingly similar across these different tasks, it is necessary to adapt the URA system to jointly accommodate these diverse purposes rather than focusing solely on information transmission \cite{Liu2022Integrated}.

In this paper, we consider a URA system that provides both sensing and communication functionalities concurrently, referred to as unsourced integrated sensing and communications (UNISAC). Specifically, we incorporate the object detection task alongside the communication task to accommodate a wide range of users. The UNISAC model serves two categories of users: communication users and sensing users. For communication users, the objective is to decode the transmitted messages. For sensing users, the objective is to detect their presence while simultaneously estimating their angles of arrival (AOA). Due to the aggregation of signals from various sensing and communication users at the receiver, the URA faces the complex task of analyzing each user's signal among the interference from other users. In this paper, we investigate the achievable performance of the UNISAC model, taking into account the misdetection error and mean-squared error of AOA estimation (MSEAOA) for sensing users, in addition to the decoding error of communication users. Numerical results indicate that the proposed UNISAC model exhibits superior performance compared to conventional multiple access approaches such as treat interference as noise (TIN), time division multiple access (TDMA), ALOHA, and multiple signal classification (MUSIC).

The rest of the paper is organized as follows: In Section~\ref{SystemModel}, we outline the system model. An approximate achievable result of the UNISAC model is derived in Section~\ref{SecAchievable}. Finally, before concluding the paper in Section~\ref{Conclusion}, we present numerical results in Section~\ref{sec_NumResults}.

\textit{Notation:} Lower-case and upper-case boldface letters are used for denoting a vector and a matrix, respectively; $\mathrm{diag}(\mathbf{A})$ generates a diagonal matrix by setting all off-diagonal elements of $\mathbf{A}$ to zero; the transpose and Hermitian of matrix $\mathbf{A}$ are denoted by $\mathbf{A}^T$ and $\mathbf{A}^H$, respectively; $\mathrm{det}(\mathbf{A})$ calculates the determinant of the matrix $\mathbf{A}$; $\mathbf{I}_N$ is an $N\times N$ identity matrix; $\mathrm{Re}(t)$ gives the real part of $t$; $ \|\mathbf{A}\|^2$ calculates the squared Frobenius norm of the matrix $\mathbf{A}$; the notation $\mathbf{A}\preceq \mathbf{B}$ indicates that the matrix $\mathbf{A}-\mathbf{B}$ is negative semidefinite; $\mathcal{CN}(0,b)$ denotes the circularly symmetric complex Gaussian distribution with a mean of 0 and variance $b$; $\mathbf{0}_{n,m}$ denotes a matrix of size $n\times m$ where all elements are zero; $F_{\chi^2}(t)$ is the cumulative distribution function of the chi-squared distribution with $t$ degrees of freedom and $F_{\chi^2}^{-1}(t)$ is its inverse; $\chi_T^2$ denotes a chi-squared random variable with $T$ degrees of freedom; the function $\mathbf{t}^R$ reverses the order of elements in the vector $\mathbf{t}$.

\section{System Model}
\label{SystemModel}
Consider an ISAC system where $K_{T_s}$ sensing users (objects) and $K_{T_c}$ communication users are connected to a BS. As shown in Fig. \ref{Fig_configuratin}, the BS is equipped with an $M$-element uniform linear array (ULA) with element spacing of $d=0.5\lambda$, where $\lambda$ represents the wavelength of the received signal. Due to sporadic behaviour, only a small fraction of sensing and communication users are active at each transmission, and a significant portion of users do not transmit any signal. All active sensing and communication users simultaneously share the same length-$n$ frame for transmitting signals. To generate its transmitted signal, each communication user maps its bit sequence $\mathbf{w}_j\in \{0, 1\}^{ B_c}$ to the first $2^{B_c}$ rows of the codebook $\mathbf{A}\in \mathbb{C}^{(2^{B_s}+2^{B_c})\times n}$, and each sensing user generates its transmitted signal by randomly choosing a row among the last $2^{B_s}$ rows of the codebook $\mathbf{A}$. Note that since there are no message sequences to be transmitted by the sensing users, the value of $B_s$ lacks a predefined value and must be optimally chosen according to the system configuration.

Considering the synchronous transmission of all the users and $d=0.5\lambda$, the received signal at the BS is written as
\begin{align}
    \mathbf{Y}=\sum_{\text{\tiny $j\in \{\mathcal{A}_c,\mathcal{A}_s\}$}}  \mathbf{b}_{\theta_{j}}\mathbf{a}_j+\mathbf{Z}    \in \mathbb{C}^{M\times n}  ,\label{receivedSignal}  
\end{align}
with
\begin{align}
\mathbf{b}_{\theta_j} = \left[1 , e^{-j\pi  d\theta_j/\lambda}, e^{-j2\pi  \theta_jd/\lambda} , \cdots ,
e^{ -j\pi d(M-1)\theta_j/\lambda}\right]^T ,\label{steeringVec}    
\end{align}
where $\mathbf{a}_j\in \mathbb{C}^{ n}$ denotes the $j$th row of the matrix $\mathbf{A}$, $\theta_j$ represents the normalized AOA corresponding to the codeword $\mathbf{a}_j$ \footnote{The normalized AOA is defined as the cosine of the actual AOA, valid in the range of $-1$ to $+1$}, ${\mathcal{A}_c}\subset \{1,2,...,2^{B_c}\}$ and ${\mathcal{A}_s}\subset \{2^{B_c}+1,...,(2^{B_s}+2^{B_c})\}$ are the sets of active row indices of the codebook $\mathbf{A}$ associated with communication and sensing users, respectively (active row index refers to a row that is selected and transmitted by an active user), and $\mathbf{Z}$ is the additive Gaussian noise matrix, with each entry following $\mathcal{CN}(0, \sigma_z^2)$. Note that in the rest of the paper, the normalized AOA will simply be referred to as AOA. The objectives of the receiver are: 1) to detect the presence of active sensing users (those with indices $j\in \mathcal{A}_s$) by detecting their transmitted signals as well as estimating their corresponding AOAs, and 2) to decode the transmitted bit-sequences of active communication users ($\mathbf{w}_{j}$ for $j\in {\mathcal{A}_c}$). 

For measuring the detection and decoding performance of the UNISAC model, the per-user probability of error (PUPE) is adopted as
    \begin{align}
    \epsilon_{} &= \dfrac{\mathbb{E}\left\{{|\mathcal{L}_{c,md}|+|\mathcal{L}_{s,md}|}+ {|\mathcal{L}_{c,coll}|+|\mathcal{L}_{s,coll}|}\right\}}{{| {\mathcal{A}_c}|+| {\mathcal{A}_s}|}}\label{Eq4MARCH7}
    \end{align}
where $\mathcal{L}_{c,md}$ and $\mathcal{L}_{s,md}$ denote the list of active communication and sensing user indices that are not detected, respectively, and $\mathcal{L}_{c,coll}$ and $\mathcal{L}_{s,coll}$ are the list of communication and sensing user indices that are in collision, respectively. Moreover, concerning the AOA estimation for the sensing users, the performance metric is the MSEAOA for the successfully detected sensing users, defined as follows:
\begin{align}
\Delta &=  \dfrac{1}{|\mathcal{L}_{s,d}|}\sum_{i\in \mathcal{L}_{s,d}}^{ }\mathbb{E}\left\{ |\theta_{i}-\hat{\theta}_{i} |^2\right\},\label{Eq3_14MARCH}
\end{align}
where $\mathcal{L}_{s,d}$ represents the list of active sensing user indices that are successfully detected, and $\hat{\theta}_{i}$ is the estimate of $\theta_{i}$. 
\begin{figure}
\centering
\includegraphics[width=1\linewidth]{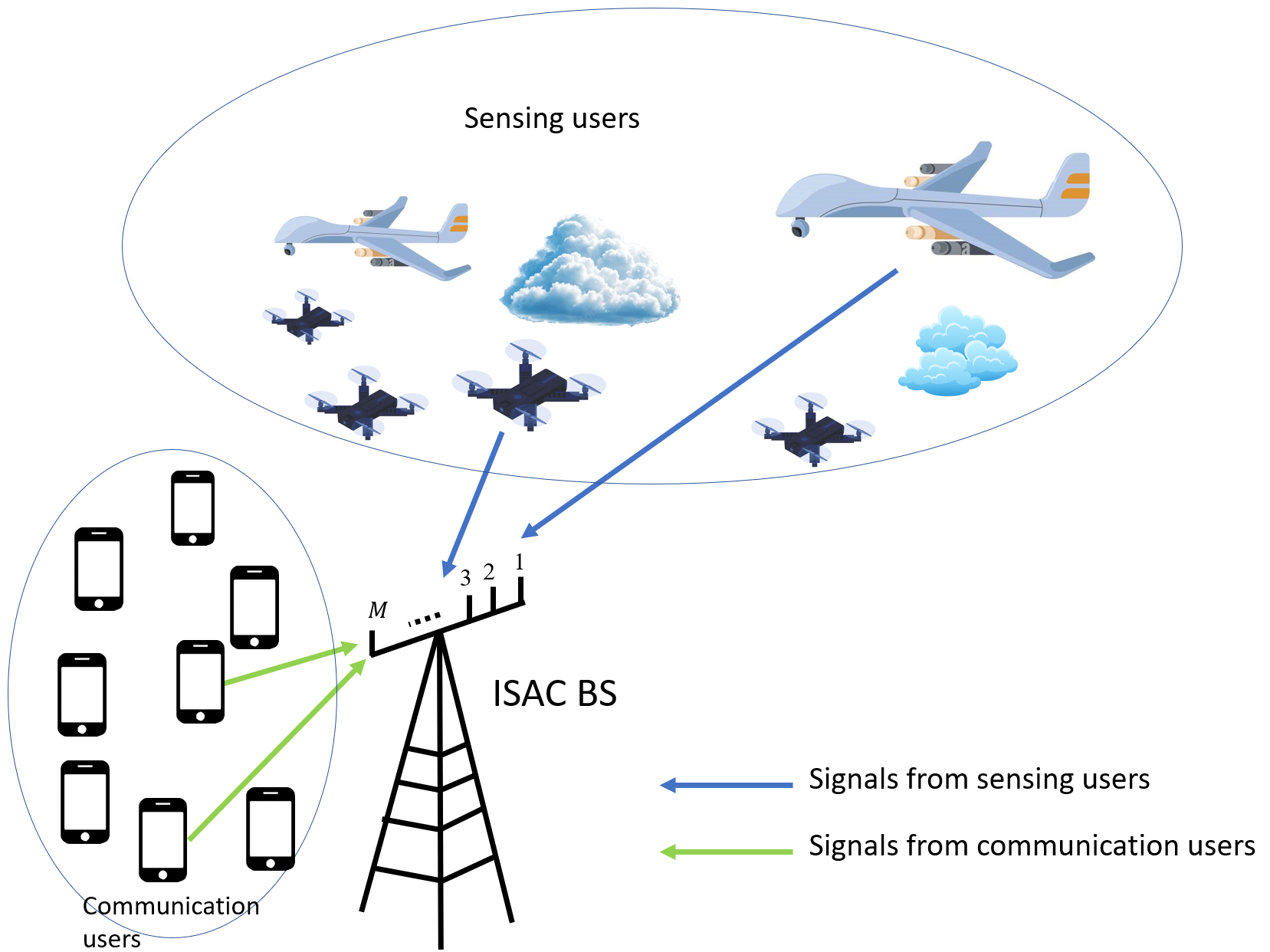}
\caption{{\small Illustration of the UNISAC model. The system accommodates multiple users for sensing and communication purposes, with only a limited subset active at any given moment.}}
    \label{Fig_configuratin}
    \end{figure}
    
    \section{Approximate Achievable Result}
\label{SecAchievable}
In this section, we derive approximate achievable bounds for the PUPE and MSEAOA of the UNISAC model.
\begin{mydef}
\label{proposition1}
For the UNISAC model, there exists a transceiver configuration where the communication and sensing users satisfy the power constraints $ \bar{P}_c$ and $\bar{P}_s$, respectively, and the performance metrics in \eqref{Eq4MARCH7} and \eqref{Eq3_14MARCH} satisfy
\begin{align}
    \epsilon_{} &\leq P_{cons}+P_{coll}+P_{md}, \label{PUPE}\\
  \Delta&\leq \sum_{K_s=0}^{|\mathcal{A}_s|}  \sum_{K_c=0}^{|\mathcal{A}_c|}P_{K_s,K_c}\mathbb{E}_{o_g}\left\{\left|\mathrm{Re}\left(\dfrac{\log(o_g)}{j\pi (M-1)} \right)\right|^2\right\},\label{MSE_eq}
\end{align}
 where $P_{cons}$ represents the probability that at least one communication/sensing user surpasses the power constraints, $P_{coll}=\dfrac{\mathbb{E}\left\{ {|\mathcal{L}_{c,coll}|+|\mathcal{L}_{s,coll}|}\right\}}{{| {\mathcal{A}_c}|+| {\mathcal{A}_s}|}}$, $P_{md}=\dfrac{\mathbb{E}\left\{{|\mathcal{L}_{c,md}|+|\mathcal{L}_{s,md}|}\right\}}{{| {\mathcal{A}_c}|+| {\mathcal{A}_s}|}}$, and $P_{K_s,K_c} = \mathbb{P}\left(|\mathcal{L}_{c,md}|=K_c,|\mathcal{L}_{s,md}|=K_s\right)$.  These values are obtained as:
  \begin{subequations}
\begin{align}
P_{cons}&=1-F_{\chi^2}\left({\dfrac{2n\bar{P}_s}{P_s^\prime}},2n\right)^{|{\mathcal{A}_s}|}F_{\chi^2}\left({\dfrac{2n\bar{P}_c}{P_c^\prime}},2n\right)^{|{\mathcal{A}_c}|}, \label{cons}\\
    P_{coll} & \leq \dfrac{\sum_{i=2}^{\infty}\dfrac{i\binom{|\mathcal{A}_s|}{i}}{2^{B_s(i-1)}}+\sum_{j=2}^{\infty}\dfrac{j\binom{|\mathcal{A}_c|}{j}}{2^{B_c(j-1)}}}{|\mathcal{A}_c|+|\mathcal{A}_s|},\label{P_col}\\
P_{md} &= \sum_{K_s=0}^{|\mathcal{A}_s|}  \sum_{K_c=0}^{|\mathcal{A}_c|}\dfrac{K_c+K_s}{|\mathcal{A}_c|+|\mathcal{A}_s|}P_{K_s,K_c},\label{eqqq59}\\
P_{K_s,K_c}& \leq e^{L_s+L_c- nM\log\left(1+0.25 \sigma_t^2/\sigma_z^2\right) },\label{P_cs}\\
L_c&=\sum_{i=0}^{K_c-1}\log\left(\dfrac{|\mathcal{A}_c|-i}{K_c-i}\right)+\log\left(\dfrac{2^{B_c}-i}{K_c-i}\right),\label{Lc_eq}\\
L_s&=\sum_{i=0}^{K_s-1}\log\left(\dfrac{|\mathcal{A}_s|-i}{K_s-i}\right)+\log\left(\dfrac{2^{B_s}-i}{K_s-i}\right),\label{Ls_eq}\\
\sigma_t^2&=K_cP_c^\prime+K_sP_s^\prime,\\
o_g&=1+(g_2+\mathbf{g}^T\mathbf{g}^R)/M,
\end{align}
\end{subequations}
where $P_c^\prime$ and $P_s^\prime$ denote the average transmitted powers of communication and sensing users, respectively, with $P_c^\prime\leq \bar{P}_c$ and $P_s^\prime\leq \bar{P}_s$, each element of $\mathbf{g}\in \mathbb{C}^{M}$ follows $\mathcal{CN}\left(0,{(\sigma_t^2+\sigma_z^2)}/{ (fP_s^\prime) }\right)$, and $g_2\sim \mathcal{CN}\left(0,4M(\sigma_t^2+\sigma_z^2) /(fP_s^\prime) \right)$, with $f$ following the chi-squared distribution with $n$ degrees of freedom.
\end{mydef}
\begin{IEEEproof}
In our analysis, we consider a random codebook $\mathbf{A}\in \mathbb{C}^{(2^{B_s}+2^{B_c})\times n}$, where every element in the first $2^{B_c}$ rows are drawn from $\mathcal{CN}(0,P_c^\prime)$, and the last $2^{B_s}$ rows consist of $\mathcal{CN}(0,P_s^\prime)$. Each communication user randomly selects its transmitted signal from the first $2^{B_c}$ rows of the codebook $\mathbf{A}$, and each sensing user selects from the last $2^{B_s}$ rows.  Note that in order to fulfill the achievability criteria, all the assumptions outlined in this section should result in an increase in errors.  \\
\indent For obtaining $\epsilon$ as shown in \eqref{PUPE}, we consider three sources of error: $P_{cons}$, $P_{coll}$, and $P_{md}$, as defined in the statement of Proposition~\ref{proposition1}. According to the definition, we have 
 \begin{align}
     P_{cons}=1-\prod_{j\in \{\mathcal{A}_c,\mathcal{A}_s\}}  \mathbb{P}\left(\|\mathbf{a}_i\|^2/n<\bar{P}_l\right).
 \end{align}
 Considering $\frac{2}{P_l^\prime}\|\mathbf{a}_i\|^2\sim \chi_{2n}^2$ for $l\in\{c,s\}$, then we can derive $P_{cons}$ as in \eqref{cons}. We calculate $P_{coll}$ in \eqref{P_col} by considering that the probability that $j$ users select the same signal is upper bounded by $\dfrac{\binom{|\mathcal{A}_l|}{j}}{2^{B_l(j-1)}}$, for $l\in \{c,s\}$. In the rest of this part, we calculate an upper bound for $P_{md}$.\\
\indent To find active $\mathbf{a}_i$'s available in the received signal in \eqref{receivedSignal}, the decoder selects $|\mathcal{A}_c|$ signals from the the first $2^{B_c}$ rows of $\mathbf{A}$, and $|\mathcal{A}_s|$ signals from the last $2^{B_s}$ rows of it. Then, appending these signals together, the matrix $\mathbf{A}_d\in \mathbb{C}^{(|\mathcal{A}_c|+|\mathcal{A}_s|)\times n}$ is constructed. Note that there are $\binom{2^{B_s}}{|\mathcal{A}_s|}\binom{2^{B_c}}{|\mathcal{A}_c|}$ possible choices for the matrix $\mathbf{A}_d$. The receiver declares the matrix that satisfies the following strategy as the collection of detected signals from active communication and sensing users:
\begin{align}
    \hat{\mathbf{A}}_d = \argmin_{\mathbf{A}_d} \|\mathbf{Y}f_p\left(\mathbf{A}_d\right)\|^2,\label{strategy}
\end{align}
where  
\begin{align}
    f_p\left(\mathbf{A}_d\right)=\mathbf{I}_n-\mathbf{A}_d^H(\mathbf{A}_d\mathbf{A}_d^H)^{-1}\mathbf{A}_d.
\end{align}
 We can write the received signal in \eqref{receivedSignal} in the following form. 
\begin{align}
    \mathbf{Y}
    =\mathbf{B}_a\mathbf{A}_a+\mathbf{Z} \label{Received11},
\end{align}
where $\mathbf{b}_{\theta_j}$'s and $\mathbf{a}_j$'s for $j\in \{\mathcal{A}_c, \mathcal{A}_s\}$ construct the columns of $\mathbf{B}_a\in \mathbb{C}^{M\times (|\mathcal{A}_c|+|\mathcal{A}_s|)}$ and the rows of $\mathbf{A}_a\in \mathbb{C}^{(|\mathcal{A}_c|+|\mathcal{A}_s|)\times n}$, respectively. Let $\mathbf{A}_a=\begin{bmatrix} \mathbf{A}_{a,1}\\\mathbf{A}_{a,2}\end{bmatrix}$ and $\mathbf{B}_a=\begin{bmatrix} \mathbf{B}_{a,1},\mathbf{B}_{a,2}\end{bmatrix}$, where $\mathbf{B}_{a,2}\in \mathbb{C}^{M\times\left(K_c+K_s\right)}$ and $\mathbf{A}_{a,2}\in \mathbb{C}^{\left(K_c+K_s\right)\times n}$ with $K_s$ and $K_c$ being arbitrary parameters. We define $\mathbf{A}_e=\begin{bmatrix} \mathbf{A}_{a,1}\\\mathbf{A}_{e,f}\end{bmatrix}$, where the rows of $\mathbf{A}_{e,f}\in \mathbb{C}^{\left(K_c+K_s\right)\times n}$ are $K_s$ sensing signals and $K_c$ communication signals that are not sent (false signals). Therefore, if $\mathbf{A}_e$ is declared as the solution to the problem in \eqref{strategy}, the signals of $K_s+K_c$ users will be incorrectly detected. Considering the definition of $P_{K_s,K_c}$ in the statement of the Proposition~\ref{proposition1} and the strategy in \eqref{strategy}, we write
\begin{align}
\nonumber
 P_{K_s,K_c}&= \mathbb{P}\left(\bigcup_{\mathbf{A}_e\in \mathcal{A}_A}\bigcap_{\mathbf{A}_e^\prime\in \Omega}\{\zeta_{\mathbf{A}_e,\mathbf{A}_e^\prime}\}\right)\\
 &\leq |\mathcal{A}_A|\mathbb{P}\left(\zeta_{\mathbf{A}_e,\mathbf{A}_a}\right),\label{PCS}
\end{align}
where $\zeta_{\mathbf{A}_1,\mathbf{A}_2}=\{\|\mathbf{Y}f_p\left(\mathbf{A}_1\right)\|^2\leq \|\mathbf{Y}f_p\left(\mathbf{A}_2\right)\|^2\}$, $\Omega$ is the universal set, and $\mathcal{A}_A$ represents the set of all possible choices for $\mathbf{A}_e$ with 
\begin{align}
    |\mathcal{A}_A|=\binom{2^{B_s}}{K_s}\binom{|\mathcal{A}_s|}{K_s}\binom{2^{B_c}}{K_c}\binom{|\mathcal{A}_c|}{K_c}.\label{Eq_A_A}
\end{align}
In \eqref{PCS}, we use the properties $\mathbb{P}\left(\bigcup_{i\in \mathcal{S}} S_i\right) \leq  \sum_{i\in \mathcal{S}}\mathbb{P}\left(S_j\right) $, and $\mathbb{P}\left(\bigcap_{i\in \mathcal{S}} S_i\right) \leq  \mathbb{P}\left(S_j\right) $, $ j \in\mathcal{S}$~\cite{Ahmadi2024RISUMA}. Since $\mathbf{A}_af_p\left(\mathbf{A}_a\right)=\mathbf{0}_{(|\mathcal{A}_c|+|\mathcal{A}_s|),n}$, $f_p\left(\mathbf{A}_a\right) \preceq \mathbf{I}_n$, and $f_p\left(\mathbf{A}_a\right)f_p\left(\mathbf{A}_a\right)^H=f_p\left(\mathbf{A}_a\right)$, then 
\begin{align}
  \|\mathbf{Y}f_p\left(\mathbf{A}_a\right)\|^2\leq \|\mathbf{Z}\|^2.\label{eq17_MARCH17f}
\end{align}
The received signal in \eqref{Received11} can be written as
\begin{align}
    \mathbf{Y}
    =\begin{bmatrix} \mathbf{B}_{a,1},\mathbf{0}_{M,(K_s+K_c)}\end{bmatrix}\mathbf{A}_{e}+\mathbf{Z}^\prime+\mathbf{Z},\label{RecevedNew}
\end{align}
where $\mathbf{Z}^\prime=\mathbf{B}_{a,2}\mathbf{A}_{a,2}$. Under the approximation that the components of the steering vectors are mutually independent, the elements of $\mathbf{Z}^\prime$ are i.i.d. and approximately distributed as $\mathcal{CN}\left(0,\sigma_t^2\right)$, where $\sigma_t^2=K_cP_c^\prime+K_sP_s^\prime$. Since elements of $\mathbf{Z}$ and $\mathbf{Z}^\prime$ are independent of the elements of $\mathbf{A}_e$, all of them being zero mean, and assuming $n$ to be large, from the weak law of large numbers we have $\mathbf{Z}^\prime\mathbf{A}_e^H\approx \mathbf{Z}\mathbf{A}_e^H\approx \mathbf{0}_{M\times (K_c+K_s)}$. Therefore, $\mathbf{Z}^\prime f_p\left(\mathbf{A}_e\right)\approx \mathbf{Z}^\prime$ and $\mathbf{Z} f_p\left(\mathbf{A}_e\right)\approx \mathbf{Z}$. Using this approximation and the property $\mathbf{A}_ef_p\left(\mathbf{A}_e\right)=\mathbf{0}_{(|\mathcal{A}_c|+|\mathcal{A}_s|),n}$, we get
\begin{align}
    \|\mathbf{Y}f_p\left(\mathbf{A}_e\right)\|^2
    &\approx \|\mathbf{Z}+\mathbf{Z}^\prime\|^2.\label{eq15_MARCH17b}
\end{align} 
Focusing on \eqref{eq17_MARCH17f} and \eqref{eq15_MARCH17b}, and taking into account that the vectorized version of a matrix has the same Frobenius norm as the matrix itself, we have
\begin{subequations}
\begin{align}
    \mathbb{P}\left( \zeta_{\mathbf{A}_e,\mathbf{A}_a}\right)&\leq \mathbb{P}\left( \|\mathbf{z}+\mathbf{z}^{\prime}\|^2<\|\mathbf{z}\|^2 \right)\\
    &\leq \mathbb{E}\left\{e^{-\lambda_1\|\mathbf{z}+\mathbf{z}^{\prime}\|^2+\lambda_1\|\mathbf{z}\|^2} \right\}\label{15b_23MARCH}\\
    &= \mathbb{E}\left\{\dfrac{e^{\lambda_1\|\mathbf{z}\|^2}e^{\dfrac{-\lambda_1\|\mathbf{z}\|^2}{(1+\lambda_1\sigma_t^2)}}}{(1+\lambda_1\sigma_t^2)^{nM}}\right\}\label{15c_23MARCH}\\
    &= \dfrac{1}{\left(1+\lambda_1 \sigma_t^2-\lambda_1^2\sigma_z^2\sigma_t^2\right)^{nM}}\label{15e_23MARCH}\\
   &= e^{{-nM\log\left(1+0.25 \sigma_t^2/\sigma_z^2\right)} }\label{UppBounPcs}   ,
\end{align}
\end{subequations}
where $\mathbf{z}$ and $\mathbf{z}^\prime$ are the vectorized versions of $\mathbf{Z}$ and $\mathbf{Z}^\prime$, respectively. To establish \eqref{15b_23MARCH}, we utilize the Chernoff bound, which states that $\mathbb{P}(x>0)\leq \mathbb{E}(e^{\lambda_1 x})$ for any $\lambda_1>0$, and to obtain \eqref{15c_23MARCH} and \eqref{15e_23MARCH}, we use the identity 
\begin{align}
    \mathbb{E}\left\{    e^{x\|\mathbf{a}+\mathbf{b}\|^2}\right\}= \dfrac{1}{(1-x\sigma_a^2)^n}     e^{x\|\mathbf{b}\|^2/(1-x\sigma_a^2)},
\end{align}
for $x\sigma_a^2<1$, $\mathbf{a}\sim \mathcal{CN}\left(\mathbf{0}_{n,1},\sigma_a^2\mathbf{I}_n\right)$, and $\mathbf{b}\in \mathbb{C}^{ n}$. To obtain \eqref{UppBounPcs}, we exploit the optimal value of $\lambda_1$ that minimizes the expression in \eqref{15e_23MARCH}, i.e., $\lambda_1=0.5/\sigma_z^2$. Plugging \eqref{UppBounPcs} and \eqref{Eq_A_A} into \eqref{PCS}, and considering that ${L_s+L_c}=\log\left(\binom{2^{B_s}}{K_s}\binom{|\mathcal{A}_s|}{K_s}\binom{2^{B_c}}{K_c}\binom{|\mathcal{A}_c|}{K_c}\right)$, an approximate upper limit for $P_{K_s,K_c}$ is obtained as in \eqref{P_cs}, with $L_c$ and $L_s$ being defined in \eqref{Lc_eq} and \eqref{Ls_eq}, respectively. \\
\indent After detecting the sensing and communication signals, we can obtain a soft estimation of their corresponding steering vectors using least-squares (LS) estimator as $\hat{\mathbf{B}}=\mathbf{Y}\mathbf{A}_e^H(\mathbf{A}_e\mathbf{A}_e^H)^{-1}$. Considering the received signal model in \eqref{RecevedNew}, we have
\begin{align}
    \hat{\mathbf{B}}&=\begin{bmatrix} \mathbf{B}_{a,1},\mathbf{0}_{M,(K_s+K_c)}\end{bmatrix}+\tilde{\mathbf{Z}},
\end{align}
where $\tilde{\mathbf{Z}}=(\mathbf{Z}+\mathbf{Z}^\prime)\mathbf{A}_e^H\left(\mathbf{A}_e\mathbf{A}_e^H\right)^{-1}$. It is clear that every row of $\tilde{\mathbf{Z}}$ approximately follows $\mathcal{CN}\left(\mathbf{0}_{ 1,J},(\sigma_z^2+\sigma_t^2)\left(\mathbf{A}_e\mathbf{A}_e^H\right)^{-1}\right)$, where $J=|\mathcal{A}_c|+|\mathcal{A}_s|-K_s-K_c$. For every $i\in \mathcal{A}_s$, the estimation of the steering vector is obtained by selecting the corresponding column of $\hat{\mathbf{B}}$, which yields
\begin{align}
    \hat{\mathbf{b}}_{\theta_i} = \mathbf{b}_{\theta_i}+\mathbf{z}_i , \label{bHHat}
\end{align}
where $\mathbf{z}_i$ is the column of $\tilde{\mathbf{Z}}$ corresponding to the index $i$. Any positive definite matrix $\mathbf{T}$ satisfies the condition $\mathbf{T}^{-1}\preceq \mathrm{diag}\left(\mathbf{T}\right)^{-1}$. Therefore, assuming every element of $\mathbf{z}_i$ to follow $\mathcal{CN}\left(0,\dfrac{\sigma_z^2+\sigma_t^2}{\|\mathbf{a}_i\|^2}\right)$ is a pessimistic assumption, leading to a higher MSEAOA due to increased noise variance. The receiver estimates $\theta_i$ as
\begin{align}
    \hat{\theta}_i=\mathrm{Re}\left(\dfrac{1}{-j\pi(M-1)}\log\left(\dfrac{1}{M}\hat{\mathbf{b}}_{\theta_i}^T\hat{\mathbf{b}}_{\theta_i}^R\right)\right). \label{Theta_hat}
\end{align}
Plugging \eqref{bHHat} into \eqref{Theta_hat} and using the structure in \eqref{steeringVec}, we get
\begin{align}
\nonumber
    \hat{\theta}_i&=\mathrm{Re}\left(\dfrac{1}{-j\pi(M-1)}\log\left(e^{-j\pi(M-1)\theta_i}+z_i^\prime\right)\right)\\
   &=\theta_i+z_i^{\prime\prime},\label{tehtea_hat}
\end{align}
where $z_i^\prime=\dfrac{2}{M}\mathbf{b}_{\theta_i}^T\mathbf{z}_i^R+\dfrac{1}{M}\mathbf{z}_i^T\mathbf{z}_i^R$, and $z_i^{\prime\prime}=\mathrm{Re}\left(\dfrac{1}{-j\pi(M-1)}\log(1+e^{j\pi(M-1)\theta_i}z_i^\prime)\right)$. Consequently, substituting \eqref{tehtea_hat} into \eqref{Eq3_14MARCH}, an upper bound for MSEAOA can be obtained as in \eqref{MSE_eq}. Note that since random shifting of $\mathbf{z}_i$ does not change its distribution, the random variable $o_g$ in the statement of Proposition~\ref{proposition1} has the same probability distribution function as $1+e^{j\pi(M-1)\theta_i}z_i^\prime$. 
\end{IEEEproof}
\begin{mydef6}
Finding $\mathbb{E}_{o_g}\left\{.\right\}$ in \eqref{MSE_eq} becomes difficult due to the complicated probability distribution function of $o_g$. Hence, we employ the Monte Carlo method to estimate it by generating multiple realizations.
\end{mydef6}
\section{Numerical Results}
\label{sec_NumResults}
\begin{figure}
\centering
		\includegraphics[width=1\linewidth]{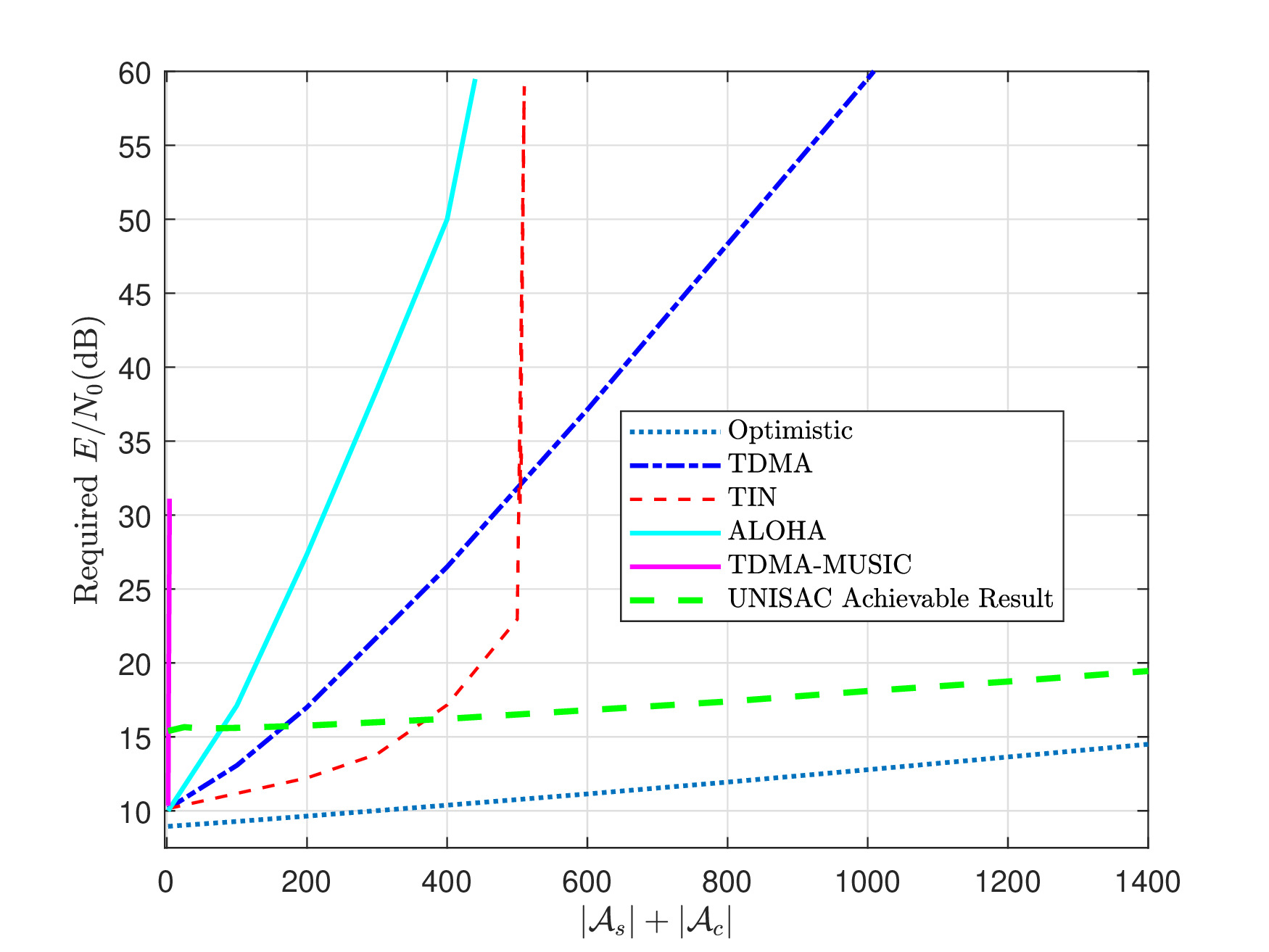}
		\caption{{\small The required $E/N_0$ as a function of the number of active sensing and communication users (with $|\mathcal{A}_c|=|\mathcal{A}_s|$), comparing UNISAC's achievable result against existing multiple access models to achieve a target PUPE of $\epsilon_0 = 0.1$ and a target MSEAOA of $\Delta_0 = 5\times 10^{-4}$. }	}
		\label{Fig_DiffModels}
	\vspace{-4. mm}
	\end{figure}

\begin{figure}
\centering
		\includegraphics[width=1\linewidth]{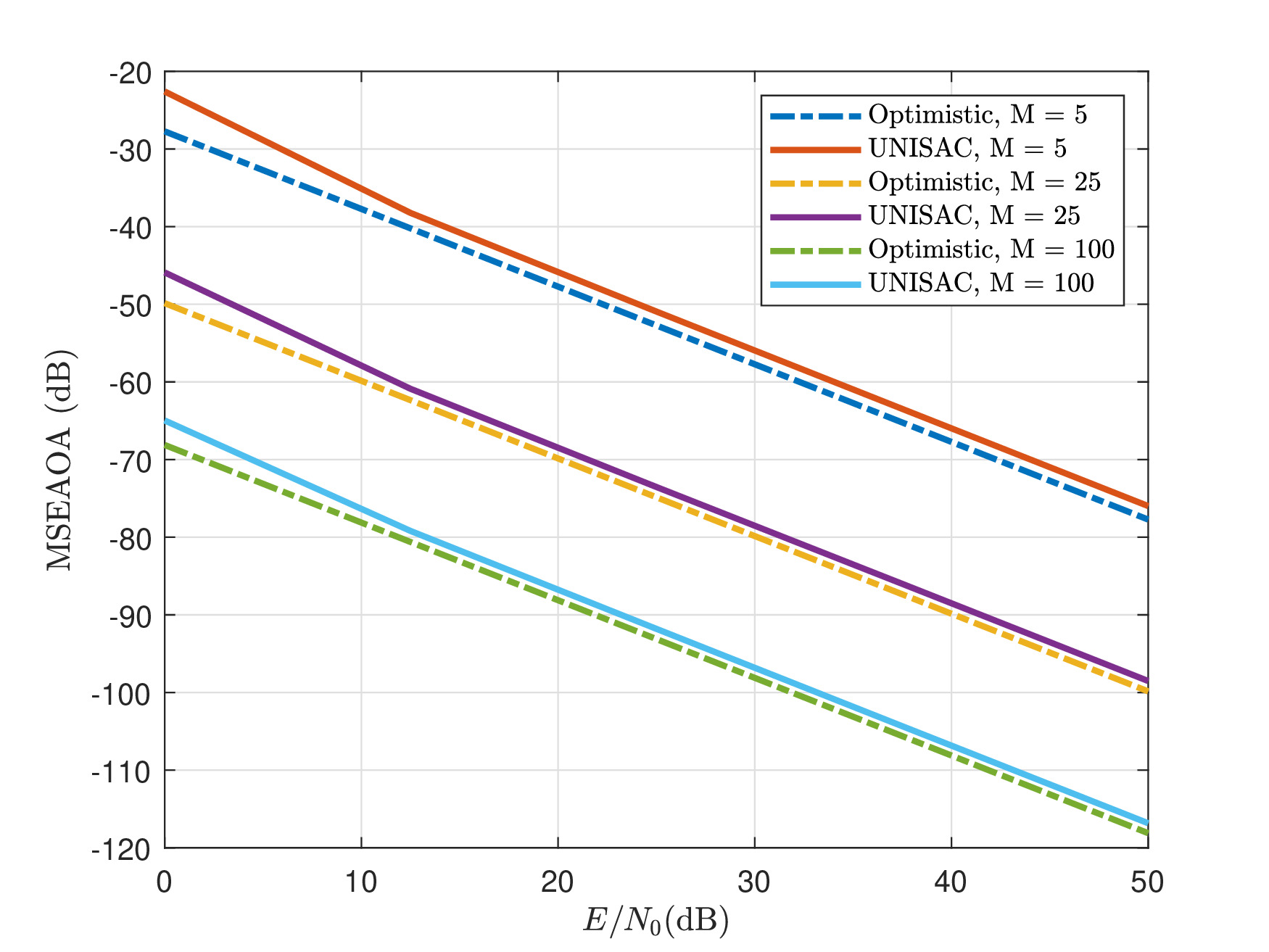}
		\caption{{\small Achievable and optimistic results of MSEAOA for different numbers of antenna elements $M$.}	}
		\label{Fig_MSEAOA}
	\end{figure}

To assess the performance of the proposed UNISAC model, we compare its approximate achievable result derived in Proposition~\ref{proposition1} with that of traditional multiple-access models. We evaluate various configurations under the following conditions: The bit-sequence length of communication users is $B_c=100$, the antenna array consists of $M=5$ elements, and the number of channel-uses is $n=5000$. Other parameters are optimized to achieve the target PUPE of $\epsilon_o=0.1$ and the target MSEAOA of $\Delta_o = 5\times 10^{-4}$, respectively.
\begin{mydef5}
\label{def1}
(\hspace{-1pt}Approximate channel coding rate in the finite blocklength regime \cite{Polyanskiy2010Channel}): For  any $P_{er} > 0$, there exists a $(2^B, n, P_{er})$ code in an additive white Gaussian noise channel, where
\begin{align}
    R\approx \log(1+S)- \sqrt{ \dfrac{1}{n}\dfrac{S(S+2)\log_2^2e}{2(S+1)^2}}Q^{-1}(P_{er}),\label{normalApprx}
\end{align}
where $R$ and $S$ denote channel coding rate and the signal-to-noise ratio (SNR), respectively, $P_{er}$ is the block error rate, $Q(.)$ denotes the standard $Q$-function, and $e$ is the Euler's number.
\end{mydef5}
\begin{mydef5}
\label{def2}
(\hspace{-1pt}\cite[Theorem 4.1]{StoicaMUSIC}): For the channel model in \eqref{Received11} with $K$ active users, the lower bound for the MSEAOA is obtained as
\begin{align}
   {  \Delta= \dfrac{\sigma_z^2}{2}\left(\sum_{t=1}^{n} \mathrm{Re}\left(\mathbf{X}_t\mathbf{D}^Hf_p(\mathbf{B}_a^H)\mathbf{D}\mathbf{X}_t^H\right) 
   \right)^{-1},} \label{CRLB}
\end{align}
where $f_p\left(\mathbf{B}\right)=\mathbf{I}_n-\mathbf{B}^H(\mathbf{B}\mathbf{B}^H)^{-1}\mathbf{B}$, $\mathbf{X}_t\in \mathbb{C}^{K\times K}$ is a diagonal matrix with the $t$th column of the matrix $\mathbf{A}_a$ on its diagonal, and $\mathbf{D}\in \mathbb{C}^{M\times K}$ represents the element-wise partial derivative of $\mathbf{B}_a$ with respect to $\theta$, $\mathbf{D}=\partial \mathbf{B}_a/\partial \theta$. 
\end{mydef5}
\begin{mydef5}
\label{def3} (\hspace{-1pt}\cite[Chapter 8]{SALEHI_Fundumental}): Let $\mathcal{H}_0$ and $\mathcal{H}_1$ be the null and alternative hypotheses, respectively, with corresponding channel models represented as follows: $\mathcal{H}_0: \mathbf{y}=\mathbf{z}$ and $\mathcal{H}_1: \mathbf{y}=\mathbf{s}+\mathbf{z}$, where $\mathbf{z}\sim \mathcal{CN}\left(\mathbf{0}_{T,1},\delta\mathbf{I}_T\right)$ represents the additive noise term and $\mathbf{s}\in \mathbb{C}^T$ is the desired signal. The misdetection probability for detecting the signal $\mathbf{s}$ can be computed as follows:
\begin{align}
    p_{miss}=Q\left(\sqrt{\dfrac{\|\mathbf{s}\|^2}{2\delta}}\right).\label{misdetection}
\end{align}
\end{mydef5}
The benchmark strategies are described below. The first model is termed the ``optimistic" model, where we adopt optimistic assumptions leading to lower bounds on PUPE and MSEAOA. The lower bound for $\mathbb{E}\left\{{|\mathcal{L}_{s,coll}|+|\mathcal{L}_{c,coll}|}\right\}$ is obtained as $\sum_{l\in\{c,s\}}\dfrac{\binom{|\mathcal{A}_l|}{2}}{2^{B_l}}\left(\dfrac{2^{B_l}-1}{{2^{B_l}}}\right)^{|\mathcal{A}_l|-2}$. To ensure that $\mathbb{E}\left\{|\mathcal{L}_{s,md}|+|\mathcal{L}_{c,md}|\right\}/({|\mathcal{A}_c|+|\mathcal{A}_s|})\xrightarrow{} 0 $, we employ Shannon limit as
\begin{align}
    \dfrac{B_T}{n} \xrightarrow{} \mathbb{E}\left\{\log_2\left(\mathrm{det}\left(\mathbf{I}_M+\dfrac{1}{\sigma_z^2}\mathbf{B}_a\Psi\mathbf{B}_a^H\right)\right)\right\}, 
\end{align}
where $B_T =|\mathcal{A}_c|B_c+|\mathcal{A}_s|B_s$ denotes the total number of transmitted bits, and $\Psi$ is a $(|\mathcal{A}_c|+|\mathcal{A}_s|)\times (|\mathcal{A}_c|+|\mathcal{A}_s|)$ diagonal matrix where the diagonal elements corresponding to the sensing users and communication devices are $\bar{P}_s$ and  $\bar{P}_c$, respectively. To obtain the lower bound of the MSEAOA, we use Result \ref{def2} for a single user to obtain 
\begin{align}
    \Delta= \dfrac{0.5\sigma_z^2}{\pi^2 n \bar{P}_s\sum_{i=1}^{M-1}i^2}.\label{MSEAOACRLB}
\end{align}
 In the second model, which is called ``TDMA" model, we divide the frame into $|\mathcal{A}_c|+|\mathcal{A}_s|$ subframes, and every active sensing and communication user transmits its signal through an individual subframe. In this case, we can obtain an approximation of the $\mathbb{E}\left\{|\mathcal{L}_{c,md}|\right\}/|\mathcal{A}_c|$ by adopting Result \ref{def1} with $S=M\bar{P}_c/\sigma_z^2$ and $R=B_c/N_{\text{\tiny TDMA}}$, where $N_{\text{\tiny TDMA}}=n/(|\mathcal{A}_c|+|\mathcal{A}_s|)$. Using \eqref{misdetection}, we obtain $\mathbb{E}\left\{|\mathcal{L}_{s,md}|\right\}/|\mathcal{A}_s|=Q\left(\sqrt{\dfrac{\bar{P}_sN_{\text{\tiny TDMA}}M}{2\sigma_z^2}}\right)$ . For MSEAOA, we employ Result \ref{def2} for a single user, which gives $\Delta = 0.5\sigma_z^2/(\pi^2 N_{\text{\tiny TDMA}} \bar{P}_s\sum_{i=1}^{M-1}i^2)$. The third model is known as ``TIN" model, where all users transmit their signals simultaneously, and the receiver treats signals of all users except a desired user as noise. In this case, $\mathbb{E}\left\{|\mathcal{L}_{c,md}|\right\}/|\mathcal{A}_c|$ is approximated by inserting $S=M\bar{P}_c/\sigma_n^2$ and  and $R=B_c/n$ into \eqref{normalApprx}, and $\mathbb{E}\left\{|\mathcal{L}_{s,md}|\right\}/|\mathcal{A}_s|\approx Q\left(\sqrt{\dfrac{\bar{P}_snM}{2\sigma_n^2}}\right)$, where $\sigma_n^2=\bar{P}_c|\mathcal{A}_c|+\bar{P}_s|\mathcal{A}_s|+\sigma_z^2$. The MSEAOA is calculated using Result \ref{def2} by employing the single-user scenario as $\Delta = 0.5\sigma_n^2/(\pi^2 n \bar{P}_s\sum_{i=1}^{M-1}i^2)$. The ``TDMA-MUSIC" model divides the length-$n$ frame into two subframes with lengths $n/2$. In the first subframe, the TDMA strategy is employed for decoding the communication users, and in the second subframe, $|\mathcal{A}_s|$ sensing users transmit their signals simultaneously. The AOAs of the sensing users in the second subframe are determined using MUSIC estimator. For the communication part, $\mathbb{E}\left\{|\mathcal{L}_{c,md}|\right\}/|\mathcal{A}_c|$ is approximated by putting $S=M\bar{P}_c/\sigma_z^2$ and $R=B_c/N_{\text{\tiny TDMA2}}$ into \eqref{normalApprx}, where $N_{\text{\tiny TDMA2}}=0.5n/|\mathcal{A}_c|$. The MSEAOA is obtained by \eqref{CRLB}, considering $|\mathcal{A}_s|$ users' transmission through $n/2$ channel uses. In the ``ALOHA" model, we divide the frame into $T$ subframes, and each user randomly selects a subframe to transmit its signal. The receiver only identifies signals from users that are not involved in collisions. The block error of the sensing and communication users is approximated by $\mathbb{E}\left\{|\mathcal{L}_{l,md}|\right\}/|\mathcal{A}_l|\approx 1-(1-P_{md,l})\left(1-f_{po}(1;\alpha_{po})/\sum_{i=1}^{\infty}f_{po}(i;\alpha_{po})\right)$, where $\alpha_{po}= (|\mathcal{A}_s|+|\mathcal{A}_c|)/T$, $l\in\{c,s\}$, $f_{po}(i;a)$ denotes the probability distribution function of the Poisson distribution with the parameter $a$, $P_{md,c}$ is approximated by plugging $S=M\bar{P}_c/\sigma_z^2$ and $R=B_c T /N_{\text{\tiny ALOHA}}$ in \eqref{normalApprx}, and $P_{md,s}$ is calculated as $P_{md,s}=Q\left(\sqrt{\dfrac{\bar{P}_sN_{\text{\tiny ALOHA}}M}{2\sigma_z^2}}\right)$, with $N_{\text{\tiny ALOHA}}=n/T$. In these computations, we use the approximation that the number of users engaged in an $i$-collision follows a Poisson distribution with a parameter of $\alpha_{po}$ \cite{Ahmadi2023Unsourced}. Also, the MSEAOA is obtained using Result \ref{def2}, which gives $ \Delta= 0.5\sigma_z^2/(\pi^2 N_{\text{\tiny ALOHA}} \bar{P}_s\sum_{i=1}^{M-1}i^2)$. \\
\indent In Fig. \ref{Fig_DiffModels}, we compare the achievable performance of the proposed UNISAC with the existing multiple access models by plotting the required ${E}/{N_0}$ to reach the target PUPE and MSEAOA, where ${E}/{N_0}$ is the average per-user energy of the system which is defined as
\begin{align}
    \dfrac{E}{N_0}=\dfrac{(\bar{P}_c |\mathcal{A}_c|n_c+\bar{P}_s |\mathcal{A}_s|n_s)}{( |\mathcal{A}_c|+ |\mathcal{A}_s|)\sigma_z^2},
\end{align}
where $n_c$ and $n_s$ denote the signal lengths of each communication and sensing user, respectively. The $x$-axis in this figure is $|\mathcal{A}_s|+|\mathcal{A}_c|$ with $|\mathcal{A}_s|=|\mathcal{A}_c|$. Observing Fig. \ref{Fig_DiffModels}, it is clear that the UNISAC model demonstrates significantly better performance than other models under conditions with a high number of active users. The TDMA-MUSIC model's underperformance stems from its inability to handle cases where the number of sensing users surpasses the number of antenna elements. The poor performance of ALOHA, TIN, and TDMA with a large number of users are attributed to dense collisions, high interference levels, and short signal lengths, respectively. Conversely, the UNISAC model has potential to achieve superior performance by mitigating collisions, reducing interference, and assigning longer signal lengths.\\
 In order to evaluate the effectiveness of the achievable result for the MSEAOA, Fig. \ref{Fig_MSEAOA} compares the MSEAOA presented in \eqref{MSE_eq} with the optimistic outcome illustrated in \eqref{MSEAOACRLB}, considering various values of $M$. To obtain this result, we assume that $P_{K_s,K_c} = 1$ when $K_s=0$ and $K_c=0$, and $P_{K_s,K_c} = 0$ otherwise. This figure illustrates the enhancement in the AOA estimation achieved by the UNISAC model as the number of antenna elements is increased. Additionally, the minimal difference between the achievable and optimistic results indicates a tight capacity region for the AOA estimation provided by the UNISAC model.
 \vspace{-5. mm}
\section{Conclusions}
\label{Conclusion}
In this paper, we have addressed the problem of ISAC in scenarios with a massive number of unsourced users. The proposed system allows active communication and sensing users to share a short transmission frame without coordination with the base station. We introduced an approximate achievable performance bound for the proposed ISAC system and compared its effectiveness against traditional models like ALOHA, TDMA, TIN, and MUSIC. Through computer simulations, we have verified the effective performance of the proposed model in detecting and decoding a large number of users. This research contributes to advancing the understanding and practical implementation of ISAC systems in real-world scenarios with massive user counts.


\begin{thebibliography}{00}
\bibitem{polyanskiy2017perspective} Y.~Polyanskiy, ``A perspective on massive random-access,'' in \emph{Proc. IEEE Int. Symp. Inf. Theory (ISIT)}, Aachen, Germany,  June 2017, pp. 2523--2527.
\bibitem{ahmadi2021random} M. J. Ahmadi and T. M. Duman, ``Random spreading for unsourced MAC with power diversity,'' \emph{IEEE Commun. Lett.}, vol. 25, no. 12, pp. 3995--3999, Dec. 2021.
\bibitem{Zhang2025SparseCode} Z. Zhang \emph{et al.}, ``Sparse code transceiver design for unsourced random access with analytical power division in Gaussian MAC,” in {\it Proc. IEEE Veh. Technol. Conf. (VTC)}, Chengdu, China, Oct. 2025, pp. 1-5.
\bibitem{ahmadi2021Unsourced}  M. J. Ahmadi and T. M. Duman, ``Unsourced random access with a massive MIMO receiver using multiple stages of orthogonal pilots,'' in \emph{Proc. IEEE Int. Symp. Inf. Theory (ISIT)}, Espoo, Finland, July 2022, pp. 2880-2885.
\bibitem{Gkagkos2023FASURA} M. Gkagkos, K. R. Narayanan, J. F. Chamberland and C. N. Georghiades, ''FASURA: A scheme for quasi-static fading unsourced random access channels,'' \emph{IEEE Trans. Commun.}, July 2023.
\bibitem{Fengler2023Coded} A. Fengler, A. Lancho and Y. Polyanskiy, ``Coded orthogonal modulation for the multi-antenna MAC,'' in \emph{Proc. IEEE 12th Int. Symp. Topics in Coding (ISTC)}, Brest, France, Sep. 2023, pp. 1--5.
\bibitem{glebov2019achievability} A. Glebov, N. Matveev, K. Andreev, A. Frolov and A. Turlikov, ``Achievability bounds for T-fold irregular repetition slotted ALOHA scheme in the Gaussian MAC,'' in \emph{Proc. IEEE Wireless Commun. Netw. Conf. (WCNC)}, Marrakesh, Morocco, Apr. 2019, pp. 1--6.
\bibitem{pradhan2020polar} A.~K. Pradhan, V.~K. Amalladinne, K.~R. Narayanan, and J.-F. Chamberland, ``Polar coding and random spreading for unsourced multiple access,'' in \emph{Proc. IEEE Int. Conf. Commun. (ICC)}, Dublin, Ireland, June 2020, pp. 1--6.
\bibitem{andreev2020polar} K. Andreev, E. Marshakov and A. Frolov, ``A polar code based TIN-SIC scheme for the unsourced random access in the quasi-static fading MAC,'' in \emph{Proc. IEEE Int. Symp. Inf. Theory (ISIT)}, Los Angeles, USA, June 2020, pp. 3019--3024.
\bibitem{Ozates2023Slotted} M. Ozates, M. Kazemi and T. M. Duman, ``A slotted pilot-based unsourced random access scheme with a multiple-antenna receiver,'' \emph{IEEE Trans. Wireless Commun.}, vol. 23, no. 4, pp. 3437--33449, April 2024.
\bibitem{Ahmadi_PHD_thesis} M. J. Ahmadi, ``Novel unsourced random access algorithms over Gaussian and fading channels,'' Ph.D. dissertation, Bilkent university, 2024.
\bibitem{Ahmadi2023RIS}  M. J. Ahmadi, M. Kazemi, and T. M. Duman, ``RIS-aided unsourced random access,'' in \emph{Proc. IEEE Global Commun. Conf. (GLOBECOM)}, Kuala Lumpur, Malaysia, Feb. 2024, pp. 3270--3275.
\bibitem{Ahmadi2024RISUMA} M. J. Ahmadi, M. Kazemi, and T. M. Duman, ``RIS-aided unsourced multiple access (RISUMA): Coding strategy and performance limits,'' {\it IEEE Trans. Wireless Commun.}, vol. 24, no. 7, pp. 6225--6239, Jul. 2025.
\bibitem{Ahmadi2025Feedback} M. J. Ahmadi \emph{et al.}, ``Efficient feedback design for unsourced random access with integrated sensing and communication,'' {\it IEEE Wireless Commun. Lett.}, vol. 15, pp. 710-714, 2026.

 \bibitem{Qi2022integrating} Q. Qi, X. Chen, A. Khalili, C. Zhong, Z. Zhang, and D. W. K. Ng,``Integrating sensing, computing, and communication in 6G wireless networks: Design and optimization,'' \emph{IEEE Trans. Commun.}, vol. 70, no. 9, pp. 6212--6227, Sep. 2022.

 \bibitem{Xu2022Anti} J. Xu, D. Li, Z. Zhu, Z. Yang, N. Zhao and D. Niyato, ``Anti-jamming design for integrated sensing and communication via aerial IRS," \emph{IEEE Trans. Commun.},'' \emph{IEEE Trans. Commun.}, vol. 72, no. 8, pp. 4607--4619, Aug. 2024.


 \bibitem{Liu2022Integrated} F. Liu, Y. Cui, C. Masouros, J. Xu, T. X. Han, Y. C. Eldar, and S. Buzzi, ``Integrated sensing and communications: Toward dual-functional wireless networks for 6G and beyond,'' \emph{IEEE J. Sel. Areas Commun.}, vol. 40, no. 6, pp. 1728--1767, June 2022.


 \bibitem{Polyanskiy2010Channel} Y. Polyanskiy, H. V. Poor and S. Verdú, ``Channel coding rate in the finite blocklength regime,'' \emph{IEEE Trans. Inf. Theory}, vol. 56, no. 5, pp. 2307--2359, May 2010.
\bibitem{StoicaMUSIC} P. Stoica and A. Nehorai, “MUSIC, maximum likelihood, and Cramer-Rao bound,'' \emph{IEEE Trans. Acoust., Speech, Signal Process.}, vol. 38, no. 12, pp. 2140--2150, Dec 1990.
\bibitem{SALEHI_Fundumental} J. Proakis and M. Salehi, \emph{Digital Communications,} 5th ed. Boston, MA, USA: McGraw-Hill Educ., 2007.

\bibitem{Ahmadi2023Unsourced}  M. J. Ahmadi, M. Kazemi, and T. M. Duman, ``Unsourced random access with a massive MIMO receiver using multiple stages of orthogonal pilots: MIMO and single-antenna structures,'' \emph{IEEE Trans. Wireless Commun.}, vol. 23, no. 2, pp. 1343--1355, Feb. 2024.
\bibitem{Abadi2019Diffusion} M. S. E. Abadi \emph{et al.}, ``Diffusion improved multiband-structured subband adaptive filter algorithm with dynamic selection of nodes over distributed networks,'' \emph{IEEE Trans. Circuits Syst. II, Exp. Briefs,}, vol. 66, no. 3 , pp. 507-511, Mar. 2019.



























 


























































\balance

\end{thebibliography}
\end{document}